# Nonlinear PI current control of reluctance synchronous machines

C.M. Hackl[⋆,†], M.J. Kamper[‡], J. Kullick[†] and J. Mitchell[‡]

***Abstract*—This paper discusses nonlinear proportional-integral (PI) current control with anti-windup of reluctance synchronous machines (RSMs) for which the flux linkage maps are known. The nonlinear controller design is based on the tuning rule "Magnitude Optimum criterion" [1]. Due to the nonlinear flux linkage, the current dynamics of RSMs are highly nonlinear and, so, the parameters of the PI controllers and the disturbance compensation feedforward control must be adjusted online (e.g., for each sampling instant). The theoretical results are illustrated and validated by simulation *and* measurement results.**

## Notation

$\mathbb{N}, \mathbb{R}$: natural, real numbers. $\boldsymbol{x} := (x_1, \ldots, x_n)^\top \in \mathbb{R}^n$: column vector, $n \in \mathbb{N}$. $\mathbf{0}_n \in \mathbb{R}^n$: zero vector. $\|\boldsymbol{x}\| := \sqrt{\boldsymbol{x}^\top \boldsymbol{x}}$: Euclidean norm of $\boldsymbol{x}$. $\boldsymbol{A} \in \mathbb{R}^{n\times m}$: real matrix, $n, m \in \mathbb{N}$, $\det(\boldsymbol{A})$: determinant of $\boldsymbol{A}$. $\boldsymbol{I}_n \in \mathbb{R}^{n\times n}$: identity matrix. $\mathcal{C}^1(I; Y)$: space of continuously differentiable functions mapping $I \to Y$. For modeling of electrical machines, a signal $\boldsymbol{\xi}$ may be represented in the three-phase $(a, b, c)$-reference frame $\boldsymbol{\xi}^{abc} := \left(\xi^a,\ \xi^b,\ \xi^c\right)^\top$, the stator-fixed $s$-reference frame $\boldsymbol{\xi}^s := \left(\xi^\alpha,\ \xi^\beta\right)^\top$ and the arbitrarily rotating $k$-reference frame $\boldsymbol{\xi}^k := \left(\xi^d,\ \xi^q\right)^\top$, which are related by $\boldsymbol{\xi}^k = \boldsymbol{T}_\mathrm{p}(\phi_\mathrm{k})^{-1}\boldsymbol{\xi}^s = \boldsymbol{T}_\mathrm{p}(\phi_\mathrm{k})^{-1}\boldsymbol{T}_\mathrm{c}\boldsymbol{\xi}^{abc}$. $\phi_\mathrm{k}$ [rad] is the (electrical) angle of the $k$-reference frame with respect to the $s$-reference frame and $\boldsymbol{T}_\mathrm{p}(\phi_\mathrm{k}) = \left[\begin{smallmatrix}\cos(\phi_\mathrm{k}) & -\sin(\phi_\mathrm{k}) \\ \sin(\phi_\mathrm{k}) & \cos(\phi_\mathrm{k})\end{smallmatrix}\right]$, $\boldsymbol{J} = \left[\begin{smallmatrix}0 & -1 \\ 1 & 0\end{smallmatrix}\right]$ and $\boldsymbol{T}_\mathrm{c} = \frac{2}{3}\left[\begin{smallmatrix}1 & -\frac{1}{2} & -\frac{1}{2} \\ 0 & \frac{\sqrt{3}}{2} & -\frac{\sqrt{3}}{2}\end{smallmatrix}\right]$ are Park, rotation (by $\frac{\pi}{2}$) and (amplitude correct) Clarke transformation matrix, respectively (see [2, 3]).

## I. Introduction

Synchronous machines are widely-spread actuators in industry due to their compact design, high efficiency and reliability. In particular, the reluctance synchronous machine (RSM) might be a viable alternative to AC drives [4, 5]: Compared to permanent-magnet synchronous machines, the RSM is simple to manufacture (e.g., for transversally laminated RSMs, the rotor consists of punched and glued iron sheets [6]) and cheap (e.g., rare earth are not necessary [7]). Moreover, due to their anisotropic flux linkage, saliency-based encoderless control schemes are applicable [8, 9], which might promote the use of RSMs in the future. However, in most cases, for the operation of synchronous machines, an inverter, parameter estimation and adequate control are required. A method for parameter estimation of RSMs with constant inductances (linear machine) at standstill is presented in [10].

In general, the RSM is highly nonlinear (in particular, in the flux linkage) and, therefore, RSMs are not easy to control with standard control methods (e.g., classical field oriented control [11, Cha. 16]). The (negative) effects of magnetic cross saturation on the control performance have been investigated in [12]. The highly nonlinear relation between stator currents and flux linkages, caused by magnetic saturation and cross magnetization effects, complicates the system description and controller design for RSMs. Moreover, the machines "inductances" are severely affected by the rotor dimensions and the resulting cross-magnetization [13]. Often, see e.g. [14–18], the flux linkage is described as a (matrix) product of (possibly nonlinear) inductances and currents, i.e., $\boldsymbol{\psi}_\mathrm{s}^k(\boldsymbol{i}_\mathrm{s}^k) = \boldsymbol{L}_\mathrm{s}^k(\boldsymbol{i}_\mathrm{s}^k)\boldsymbol{i}_\mathrm{s}^k$. However, this kind of modeling has several disadvantages: (a) It cannot reproduce non-zero flux linkages for zero currents, i.e. $\boldsymbol{i}_\mathrm{s}^k = \mathbf{0}_2$, since $\boldsymbol{\psi}_\mathrm{s}^k(\mathbf{0}_2) = \mathbf{0}_2$, (b) it may lead to a lower control bandwidth or even to instability [19], and (c) its time derivative $\frac{\mathrm{d}}{\mathrm{d}t}\boldsymbol{\psi}_\mathrm{s}^k(\boldsymbol{i}_\mathrm{s}^k) = \boldsymbol{L}_\mathrm{s}^k(\boldsymbol{i}_\mathrm{s}^k)\boldsymbol{i}_\mathrm{s}^k$ results in *stationary* and *transient* inductances (which have no physical counterpart and are mathematically questionable [20]).

Regarding the control of RSMs, two main ideas have been subject to extensive research in the past years: Direct Torque Control (DTC), as first proposed by Boldea [21] and Lagerquist [22] as Torque Vector Control (TVC) and (classical) vector control as discussed e.g. in [23–25]. While DTC is known for its robustness and fast dynamics [26], it produces a high current distortion leading to torque ripples [27]. In contrast, vector control improves the torque response [28] and the efficiency of the system [7], but good knowledge of the system parameters is required for implementation. In [29], a completely parameter-free adaptive PI controller is proposed which guarantees tracking with prescribed transient accuracy. The controller is applied to current control of (reluctance) synchronous machines but measurement results are not provided.

Other control approaches (see [28] and [17]) track the inductances online to adjust the current *references* to achieve a higher control accuracy. In [14, 30] a control scheme is proposed, where the PI control parameters are continuously adapted to the actual system state. This improves the overall current dynamics. However, the described system model requires measurement or estimation of the machine's stationary and transient inductances (which are not easy to measure since both are *not* physical quantities) and the implementation is based on an approximation of the measured data (as the data

*Corresponding author, e-mail: christoph.hackl@tum.de (All authors, in alphabetical order, contributed equally to this paper).

†C.M. Hackl and J. Kullick are with the research group "Control of renewable energy systems" (CRES) at the Munich School of Engineering (MSE), Technische Universität München (TUM), Germany.

‡M.J. Kamper and J. Mitchell are with the Electric Machines Laboratory (EMLab) at the University of Stellenbosch, South Africa.

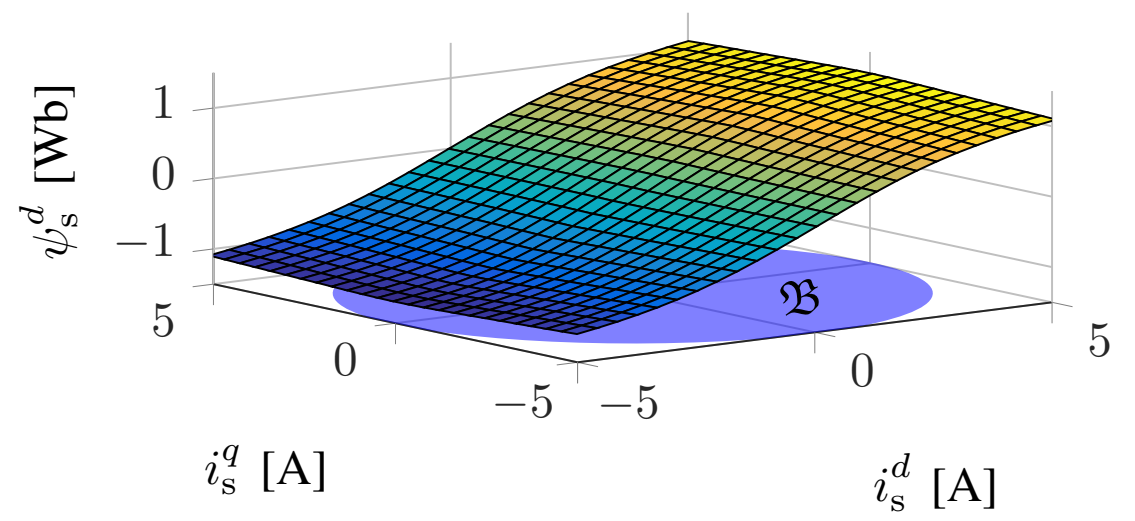

(a) $d$-component $\psi_s^d$ of the stator flux linkage.

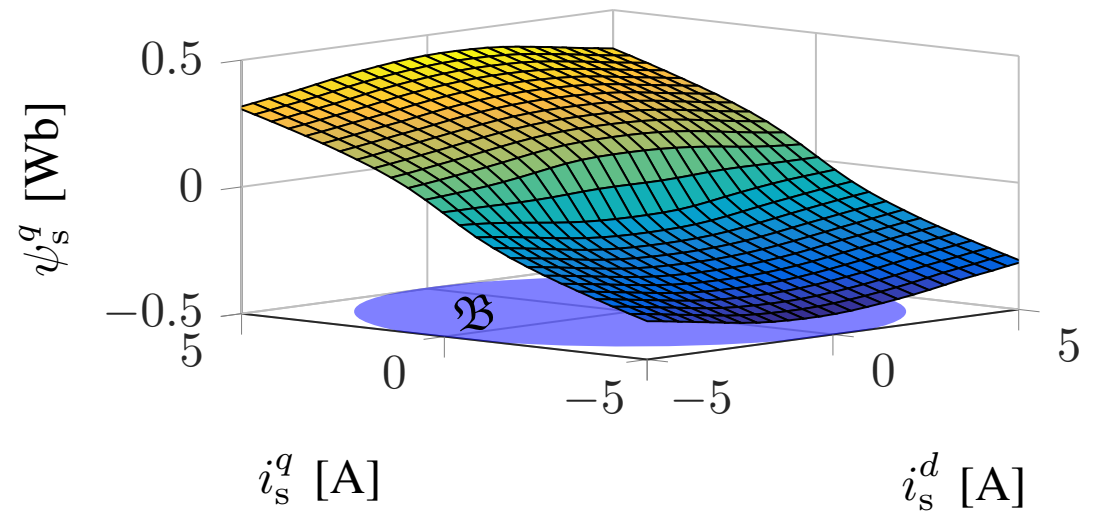

(b) $q$-component $\psi_s^q$ of the stator flux linkage.

Fig. 1: Nonlinear flux linkage of a real RSM (obtained from FEM data).

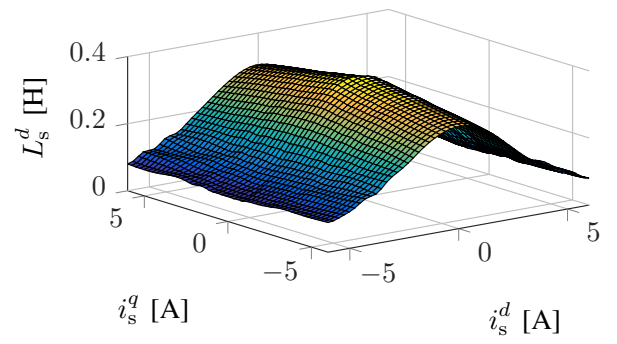

(a) $d$-inductance $L_s^d := \partial\psi_s^d/\partial i_s^d$.

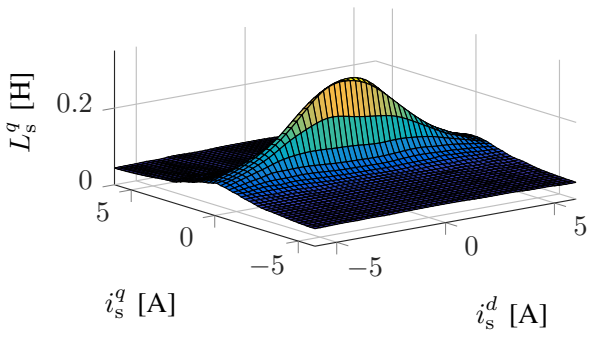

(b) $q$-inductance $L_s^q := \partial\psi_s^q/\partial i_s^q$.

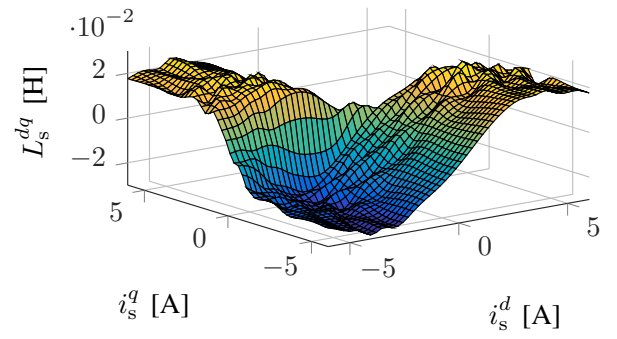

(c) cross-coupling inductance $L_s^{dq} := \partial\psi_s^d/\partial i_s^q$.

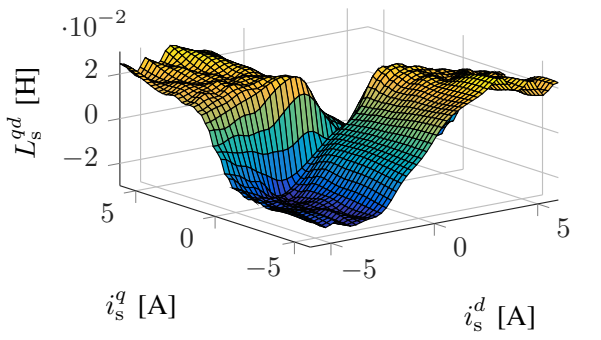

(d) cross-coupling inductance $L_s^{qd} := \partial\psi_s^q/\partial i_s^d$.

Fig. 2: Nonlinear *differential* inductances of a 1,1 kW RSM (computed from the FEM flux linkage data shown in Fig. 1).

capacity is limited on the test system). In [15], a predictive torque controller is proposed which takes into account the magnetic cross saturation of the RSM.

*Contribution of this paper:* In this paper, it is assumed that the nonlinear flux linkage maps are known (e.g. as look-up tables). Such look-up tables can easily be computed from the FEM data and, if 2D interpolation is used, their size "can be fairly small" [31]. Hence, the nonlinear relation between stator currents and flux linkages is known and can be visualized as shown in Fig. 1. Moreover, the *differential* inductances (see Fig. 2) may be approximated by numerical interpolation and differentiation of the flux map with respect to the stator currents. The look-up tables for the flux linkages and the differential inductances will be used for online adjustment of (i) the current PI controllers tuned according to the "Magnitude Optimum criterion" [1] and (ii) the disturbance compensation (feedforward control). The online adjustment guarantees an almost identical closed-loop system response of the current dynamics over the whole operation range. Main contributions of this paper are: (i) generic model of RSMs and the derivation of the current dynamics with *differential* inductances (see Sec. II-A; iron losses or hysteresis effects are *not* considered), (ii) generic model of two-level voltage source inverter *without* imposing the assumption of balanced phase voltages (see Sec. II-B), (iii) theoretical derivation of the proposed nonlinear PI controller design with anti-windup according to the "Magnitude Optimum criterion" with online adjustment of the controller parameters and of the disturbance compensation (see Sec. III), and (iv) validation of the proposed approach (modeling, controller design, and implementation) and illustration of the control performance of the nonlinear PI controllers and the nonlinear disturbance compensation by simulation and measurement results (see Sec. IV).

## II. Modeling and problem formulation

RSMs are considered which are subject to unknown disturbances (e.g. load torques). Friction is neglected. The RSM is actuated by a two-level voltage source inverter (VSI) with a *constant* DC-link voltage. Control objectives is stable, fast and accurate current reference tracking of some bounded reference (e.g., provided by a maximum-torque-per-Ampere algorithm) with (almost) identical closed-loop current dynamics over the whole operation range of the RSM such that the design of outer loop controllers remains simple.

### A. *Generic model of reluctance synchronous machine (RSM)*

The machine model in the rotating $k$-reference frame is given by (see, e.g., [11, Sec. 16.1])

$$\left.\begin{aligned} \boldsymbol{u}_s^k &= R_s\boldsymbol{i}_s^k + \omega_k \boldsymbol{J}\boldsymbol{\psi}_s^k(\boldsymbol{i}_s^k) + \tfrac{\mathrm{d}}{\mathrm{d}t}\boldsymbol{\psi}_s^k(\boldsymbol{i}_s^k), \\ \tfrac{\mathrm{d}}{\mathrm{d}t}\omega_k &= \frac{p}{\Theta}\Big[m_m(\boldsymbol{i}_s^k) - m_l\Big], \qquad \tfrac{\mathrm{d}}{\mathrm{d}t}\phi_k = \omega_k \end{aligned}\right\} \tag{1}$$

where $\boldsymbol{u}_s^k := (u_s^d, u_s^q)^\top\,[\mathrm{V}]^2$ are the applied stator voltages (see Sec. II-B), $R_s\,[\Omega]$ is the resistance of the stator windings, $\boldsymbol{i}_s^k := (i_s^d, i_s^q)^\top\,[\mathrm{A}]^2$ are the stator currents and $\boldsymbol{\psi}_s^k := (\psi_s^d, \psi_s^q)^\top\,[\mathrm{Wb}]^2$ are the stator flux linkages (functions of $\boldsymbol{i}_s^k$). The $k$-reference frame rotates with electrical angular frequency $\omega_k = p\,\omega_m[\mathrm{rad/s}]$ of the rotor where $p\,[1]$ is the number of pole pairs and $\omega_m[\mathrm{rad/s}]$ denotes the mechanical angular frequency of the machine (for details see, e.g., [3]). Furthermore, $\Theta\,[\mathrm{kgm}^3]$ is the inertia, $\boxed{m_m(\boldsymbol{i}_s^k) := \tfrac{3}{2}p\,(\boldsymbol{i}_s^k)^\top \boldsymbol{J}\boldsymbol{\psi}_s^k(\boldsymbol{i}_s^k)}$ is the machine torque, and $m_l\,[\mathrm{Nm}]$ is a load torque.

### B. *Generic model of the voltage source inverter (VSI)*

The control input to the electrical drive system (i.e. RSM *and* VSI) in Fig. 4 is the reference stator voltage $\boldsymbol{u}_{s,\mathrm{ref}}^{abc}$. The (two-level) VSI generates the stator phase voltages $\boldsymbol{u}_s^{abc} = (u_s^a, u_s^b, u_s^c)^\top$ which are then applied to the RSM terminals.

*1) Switching model of the VSI:* The considered two-level VSI is modeled as follows

$$\boldsymbol{u}_{\rm s}^{ltl}(t) := \begin{pmatrix} u_{\rm s}^{ab}(t) \\ u_{\rm s}^{bc}(t) \\ u_{\rm s}^{ca}(t) \end{pmatrix} = u_{\rm dc} \begin{bmatrix} 1 & -1 & 0 \\ 0 & 1 & -1 \\ -1 & 0 & 1 \end{bmatrix} \boldsymbol{s}_{\rm s}^{abc}(t). \qquad (2)$$

Its output voltages are the stator line-to-line voltages $\boldsymbol{u}_{\rm s}^{ltl}$ and depend on the switching vector $\boldsymbol{s}_{\rm s}^{abc}$ and the DC-link voltage $u_{\rm dc} > 0\,[\mathrm{V}]$. The switching vector $\boldsymbol{s}_{\rm s}^{abc} \in \{000, 001, \ldots, 111\}$ comprises eight different states and is generated via pulse width modulation (PWM) or space vector modulation (SVM). Applying the Clarke transformation to the three-phase stator voltages $\boldsymbol{u}_{\rm s}^{abc}$ yields (see also [3])

$$\boldsymbol{u}_{\rm s}^{s} := \begin{pmatrix} u_{\rm s}^{\alpha} \\ u_{\rm s}^{\beta} \end{pmatrix} = \boldsymbol{T}_{\rm c}\boldsymbol{u}_{\rm s}^{abc} = \underbrace{\tfrac{2}{3}\begin{bmatrix} \frac{1}{2} & 0 & -\frac{1}{2} \\ 0 & \frac{\sqrt{3}}{2} & 0 \end{bmatrix}}_{=:\boldsymbol{T}_{\rm c}^{ltl}} \boldsymbol{u}_{\rm s}^{ltl}, \qquad (3)$$

which shows that the stator voltages $\boldsymbol{u}_{\rm s}^{s}$ can directly be computed via the line-to-line Clarke transformation matrix $\boldsymbol{T}_{\rm c}^{ltl}$ and the line-to-line voltages $\boldsymbol{u}_{\rm s}^{ltl}$. Moreover, $\alpha$- and $\beta$-component of the stator voltages $\boldsymbol{u}_{\rm s}^{s}$ do *not* depend on the zero component of the stator voltage $u_{\rm s}^{0} := \frac{\sqrt{2}}{3}(u_{\rm s}^{a} + u_{\rm s}^{b} + u_{\rm s}^{c})$ which, in general, is *non-zero*.

**Remark 1.** *For balanced stator voltages (i.e. $u_{\rm s}^{0} = 0$), the following holds [3]*

$$\boldsymbol{u}_{\rm s}^{abc}(t) := \begin{pmatrix} u_{\rm s}^{a}(t) \\ u_{\rm s}^{b}(t) \\ u_{\rm s}^{c}(t) \end{pmatrix} = \frac{u_{\rm dc}}{3} \begin{bmatrix} 2 & -1 & -1 \\ -1 & 2 & -1 \\ -1 & -1 & 2 \end{bmatrix} \boldsymbol{s}_{\rm s}^{abc}(t). \qquad (4)$$

For the remainder, it is assumed that a *regularly sampled, symmetrical* PWM[1] (for details see [32, Sec. 8.4]) is implemented. Due to this PWM, *on average*, the VSI output voltage is saturated by $\widehat{u} := \frac{u_{\rm dc}}{2} > 0\,[\mathrm{V}]$ [32, Sec. 8.4.5], i.e.

$$\|\boldsymbol{u}_{\rm s}^{k}\| = \|\boldsymbol{u}_{\rm s}^{s}\| = \|\boldsymbol{u}_{\rm s}^{abc}\| \leq \tfrac{u_{\rm dc}}{2} =: \widehat{u}. \qquad (5)$$

For e.g. SVM with over-modulation or PWM with third-harmonics injection, $\widehat{u} \leq \frac{2}{3}u_{\rm dc}$ increases [33, Sec. 8.4].

*2) Inverter delay (VSI dynamics):* In general, the voltage generation is subject to a delay $T_{\rm delay} \propto 1/f_{\rm s} > 0\,[\mathrm{s}]$, i.e. $\boldsymbol{u}_{\rm s}^{k}(t) = \boldsymbol{u}_{\rm s}^{k}(t - T_{\rm delay})$ ([34], this holds for all reference frames), which – on average – is inversely proportional to the switching frequency $f_{\rm s} > 0\,[\mathrm{Hz}]$. In the Laplace domain, the delay can be approximated by a first-order lag system

$$\boldsymbol{u}_{\rm s}^{k}(s) = \mathrm{e}^{-sT_{\rm delay}} \cdot \boldsymbol{u}_{\rm s,ref}^{k}(s) \approx \tfrac{1}{1+sT_{\rm delay}}\boldsymbol{u}_{\rm s,ref}^{k}(s) \qquad (6)$$

since, for $sT_{\rm delay} \ll 1$, the following approximation holds $\mathrm{e}^{-sT_{\rm delay}} = \big(\sum_{i=0}^{\infty} \frac{(sT_{\rm delay})^{i}}{i!}\big)^{-1} \approx \frac{1}{1+sT_{\rm delay}}$.

The VSI delay depends on the implementation of the modulation scheme (e.g. on FPGA, DSP or micro-processor) and varies within the interval $T_{\rm delay} \in \big[\frac{1}{2f_{\rm s}}, \frac{3}{2f_{\rm s}}\big]$ [34]. It can further be shown that the VSI delay leads to a coupling between the $d$- and $q$-component of the reference stator voltages which may be compensated for under certain assumptions [35]. For the remainder of this paper, however, this coupling effect will be neglected (see Assumption 3).

### C. Control objective: Current reference tracking

Control objective is stator current reference tracking of some given, possibly discontinuous but bounded current reference $\boldsymbol{i}_{\rm s,ref}^{k}$ such that asymptotic tracking is achieved for constant references. Therefore, nonlinear proportional-integral (PI) controllers and a nonlinear disturbance compensation (feedforward control) will be implemented. In view of the saturation (5) of the VSI (saturated output voltage), the PI controllers should incorporate an anti-windup strategy. Moreover, due to the nonlinear system dynamics, the controller parameters and the disturbance compensation terms will be adjusted online to achieve (almost) identical closed-loop current dynamics over the whole operation range of the RSM. Controller tuning is done according to the tuning rule of the "Magnitude Optimum criterion". The achievement of the control objective shall be assured under the following assumptions:

**Assumption 1** (Properties of the flux linkages)**.** *The flux linkages $\boldsymbol{\psi}_{\rm s}^{k}\,[\mathrm{Wb}]^2$ are continuously differential functions of the stator currents $\boldsymbol{i}_{\rm s}^{k}\,[\mathrm{A}]^2$ only*[2]*, i.e. $\boldsymbol{\psi}_{\rm s}^{k}(\cdot) \in \mathcal{C}^1(\mathbb{R}^2;\mathbb{R}^2)$*

**Assumption 2** (System knowledge and measured signals)**.** *Resistance $R_{\rm s}$ and flux linkages $\boldsymbol{\psi}_{\rm s}^{k}(\boldsymbol{i}_{\rm s}^{k})$ as depicted in Fig. 1 are known (e.g. from FEM or measurements). Moreover, stator currents $\boldsymbol{i}_{\rm s}^{abc}(t)$, mechanical angle $\phi_m(t)$ and mechanical angular speed $\omega_{\rm m}(t)$ are available for feedback.*

**Assumption 3.** *The VSI has the following properties: (i) Its dynamics are sufficiently fast (i.e., $T_{\rm delay} \ll 1\,\mathrm{s}$ is small compared to the stator dynamics of the RSM), such that (a) the approximation as first-order lag system as in* (6) *is reasonable and (b) the reference voltage coupling is negligible; (ii) Its DC-link voltage is constant, i.e. $u_{\rm dc}(t) = u_{\rm dc} > 0\,\mathrm{V}$, and (iii) Its (average) output voltage is limited, i.e.* (5) *holds true.*

## III. Current controller design

Controller design is discussed in two steps (i) current dynamics decoupling via disturbance compensation feedforward control and (ii) nonlinear PI controller design with anti-windup according to the "Magnitude Optimum criterion" (controller parameters are updated at each sampling instant). The derivation is shown in the time-continuous case, whereas implementation is done in the discrete-time case (see Sec. IV).

### A. Current dynamics

Since *model-based* current controllers will be designed, the current dynamics of the RSM are required. Due to the *nonlinear* current dynamics, the model will be derived in state space[3]. In view of Assumption 1, a *differential inductance matrix* $\boldsymbol{L}_{\rm s}^{k}\,[\mathrm{H}]^{2\times 2}$ can be introduced to represent the derivatives of the flux linkage with respect to the stator current $\boldsymbol{i}_{\rm s}^{k}$, i.e.

[1] In this paper, the injection of e.g. third harmonics is *not* considered [32, Sec. 8.4.6].

[2] The flux linkages do *not* depend on the electrical/mechanical rotor angle.

[3] An analysis in the frequency domain (transfer functions) is *not* admissible.

**Definition 1.** *The* differential inductance matrix *is given by*

$$\boldsymbol{L}_\mathrm{s}^k(\boldsymbol{i}_\mathrm{s}^k) := \begin{bmatrix} \frac{\partial \psi_\mathrm{s}^d(\boldsymbol{i}_\mathrm{s}^k)}{\partial i_\mathrm{s}^d} & \frac{\partial \psi_\mathrm{s}^d(\boldsymbol{i}_\mathrm{s}^k)}{\partial i_\mathrm{s}^q} \\ \frac{\partial \psi_\mathrm{s}^q(\boldsymbol{i}_\mathrm{s}^k)}{\partial i_\mathrm{s}^d} & \frac{\partial \psi_\mathrm{s}^q(\boldsymbol{i}_\mathrm{s}^k)}{\partial i_\mathrm{s}^q} \end{bmatrix}\Bigg|_{\boldsymbol{i}_\mathrm{s}^k} := \begin{bmatrix} L_\mathrm{s}^d(\boldsymbol{i}_\mathrm{s}^k) & L_\mathrm{s}^{dq}(\boldsymbol{i}_\mathrm{s}^k) \\ L_\mathrm{s}^{qd}(\boldsymbol{i}_\mathrm{s}^k) & L_\mathrm{s}^q(\boldsymbol{i}_\mathrm{s}^k) \end{bmatrix} \quad (7)$$

*and has the following properties for all admissible stator currents: $(p_1)$ It is* positive definite *(see [36]), i.e.* $L_\mathrm{s}^d(\boldsymbol{i}_\mathrm{s}^k) > 0$, $L_\mathrm{s}^q(\boldsymbol{i}_\mathrm{s}^k) > 0$ *and* $\boxed{\det\big(\boldsymbol{L}_\mathrm{s}^k(\boldsymbol{i}_\mathrm{s}^k)\big) := L_\mathrm{s}^d(\boldsymbol{i}_\mathrm{s}^k)L_\mathrm{s}^q(\boldsymbol{i}_\mathrm{s}^k) - M(\boldsymbol{i}_\mathrm{s}^k)^2 > 0}$ *and $(p_2)$ It is* symmetric *(see [6, 37]), i.e.* $\boldsymbol{L}_\mathrm{s}^k(\boldsymbol{i}_\mathrm{s}^k) = \boldsymbol{L}_\mathrm{s}^k(\boldsymbol{i}_\mathrm{s}^k)^\top$ *and* $\boxed{M(\boldsymbol{i}_\mathrm{s}^k) := L_\mathrm{s}^{dq}(\boldsymbol{i}_\mathrm{s}^k) = L_\mathrm{s}^{qd}(\boldsymbol{i}_\mathrm{s}^k)}$ *where* $M(\boldsymbol{i}_\mathrm{s}^k)\,[\mathrm{H}]$ *is the mutual (cross-coupling) inductance of the RSM.*

Note that, in view of Assumption 2, the differential inductance matrix can be computed numerically from the FEM data (as shown in Fig. 1). Due to numerical differentiation, the cross-coupling inductances might slightly differ (i.e. $L_\mathrm{s}^{dq} \neq L_\mathrm{s}^{qd}$, see Fig. 2). Invoking Assumption 1 and Definition 1 and computing the time derivative of the flux linkages (by applying the chain rule) leads to the following expression

$$\tfrac{\mathrm{d}}{\mathrm{d}t}\boldsymbol{\psi}_\mathrm{s}^k(\boldsymbol{i}_\mathrm{s}^k) = \tfrac{\partial \boldsymbol{\psi}_\mathrm{s}^k(\boldsymbol{i}_\mathrm{s}^k)}{\partial \boldsymbol{i}_\mathrm{s}^k}\tfrac{\mathrm{d}}{\mathrm{d}t}\boldsymbol{i}_\mathrm{s}^k \overset{(7)}{=} \boldsymbol{L}_\mathrm{s}^k(\boldsymbol{i}_\mathrm{s}^k)\tfrac{\mathrm{d}}{\mathrm{d}t}\boldsymbol{i}_\mathrm{s}^k. \quad (8)$$

Inserting (8) into (1) and solving for $\frac{\mathrm{d}}{\mathrm{d}t}\boldsymbol{i}_\mathrm{s}^k$ leads to the nonlinear current dynamics as follows

$$\tfrac{\mathrm{d}}{\mathrm{d}t}\boldsymbol{i}_\mathrm{s}^k = \boldsymbol{L}_\mathrm{s}^k(\boldsymbol{i}_\mathrm{s}^k)^{-1}\cdot\Big[\boldsymbol{u}_\mathrm{s}^k - R_\mathrm{s}\boldsymbol{i}_\mathrm{s}^k - \omega_\mathrm{k}\boldsymbol{J}\boldsymbol{\psi}_\mathrm{s}^k(\boldsymbol{i}_\mathrm{s}^k)\Big], \quad (9)$$

where the inverse of the inductance matrix[4] is given by

$$\boldsymbol{L}_\mathrm{s}^k(\boldsymbol{i}_\mathrm{s}^k)^{-1} = \frac{1}{\det(\boldsymbol{L}_\mathrm{s}^k(\boldsymbol{i}_\mathrm{s}^k))}\begin{bmatrix} L_\mathrm{s}^q(\boldsymbol{i}_\mathrm{s}^k) & -M(\boldsymbol{i}_\mathrm{s}^k) \\ -M(\boldsymbol{i}_\mathrm{s}^k) & L_\mathrm{s}^d(\boldsymbol{i}_\mathrm{s}^k)\end{bmatrix}. \quad (10)$$

**Remark 2.** *In view of properties $(p_1)$ and $(p_2)$ the inductance matrix $\boldsymbol{L}_\mathrm{s}^k(\boldsymbol{i}_\mathrm{s}^k)$ nonlinearly depends on the stator currents $\boldsymbol{i}_\mathrm{s}^k$, is* non*-singular (hence invertible) and has real and positive eigenvalues for all currents $\boldsymbol{i}_\mathrm{s}^k$ [38, Proposition 5.5.20].*

The current dynamics (9) are nonlinear *and* coupled due to the inverse of the inductance matrix (10) and due to the nonlinear term $\omega_\mathrm{k}\boldsymbol{J}\boldsymbol{\psi}_\mathrm{s}^k(\boldsymbol{i}_\mathrm{s}^k)$ of the back electro-motive (EMF). For the further derivation, it is convenient to introduce the *auxiliary* inductances (which also depend nonlinearly on $\boldsymbol{i}_\mathrm{s}^k$)

$$\widetilde{L}_s^d(\boldsymbol{i}_\mathrm{s}^k) := \tfrac{\det(\boldsymbol{L}_\mathrm{s}^k(\boldsymbol{i}_\mathrm{s}^k))}{L_\mathrm{s}^q(\boldsymbol{i}_\mathrm{s}^k)} > 0 \wedge \widetilde{L}_s^q(\boldsymbol{i}_\mathrm{s}^k) := \tfrac{\det(\boldsymbol{L}_\mathrm{s}^k(\boldsymbol{i}_\mathrm{s}^k))}{L_\mathrm{s}^d(\boldsymbol{i}_\mathrm{s}^k)} > 0 \quad (11)$$

and the "disturbance voltages"

$$\begin{aligned}\boldsymbol{u}_\mathrm{s,dist}^k(\omega_\mathrm{k},\boldsymbol{u}_\mathrm{s}^k,\boldsymbol{i}_\mathrm{s}^k) &:= \begin{pmatrix} u_\mathrm{s,dist}^d(\omega_\mathrm{k},\boldsymbol{u}_\mathrm{s}^k,\boldsymbol{i}_\mathrm{s}^k) \\ u_\mathrm{s,dist}^q(\omega_\mathrm{k},\boldsymbol{u}_\mathrm{s}^k,\boldsymbol{i}_\mathrm{s}^k)\end{pmatrix} \\ &:= \begin{pmatrix} \omega_\mathrm{k}\psi_\mathrm{s}^q(\boldsymbol{i}_\mathrm{s}^k) - \frac{M(\boldsymbol{i}_\mathrm{s}^k)}{L_\mathrm{s}^q(\boldsymbol{i}_\mathrm{s}^k)}\Big(u_\mathrm{s}^q - R_\mathrm{s}i_\mathrm{s}^q - \omega_\mathrm{k}\psi_\mathrm{s}^d(\boldsymbol{i}_\mathrm{s}^k)\Big) \\ -\omega_\mathrm{k}\psi_\mathrm{s}^d(\boldsymbol{i}_\mathrm{s}^k) - \frac{M(\boldsymbol{i}_\mathrm{s}^k)}{L_\mathrm{s}^d(\boldsymbol{i}_\mathrm{s}^k)}\Big(u_\mathrm{s}^d - R_\mathrm{s}i_\mathrm{s}^d + \omega_\mathrm{k}\psi_\mathrm{s}^q(\boldsymbol{i}_\mathrm{s}^k)\Big)\end{pmatrix}. \quad (12)\end{aligned}$$

The disturbance voltages $\boldsymbol{u}_\mathrm{s,dist}^k$ comprise all coupling terms and depend on the *applied* stator voltages $\boldsymbol{u}_\mathrm{s}^k$, the stator currents $\boldsymbol{i}_\mathrm{s}^k$ and the angular velocity $\omega_\mathrm{k}$, respectively. Inserting the newly introduced quantities above into (9) and solving for $\frac{\mathrm{d}}{\mathrm{d}t}i_\mathrm{s}^d$ and $\frac{\mathrm{d}}{\mathrm{d}t}i_\mathrm{s}^q$, allows to rewrite the current dynamics component-wise in the more compact form

$$\left.\begin{aligned} \tfrac{\mathrm{d}}{\mathrm{d}t}i_\mathrm{s}^d &= \tfrac{1}{\widetilde{L}_s^d(\boldsymbol{i}_\mathrm{s}^k)}\Big(u_\mathrm{s}^d - R_\mathrm{s}i_\mathrm{s}^d + u_\mathrm{s,dist}^d(\omega_\mathrm{k},\boldsymbol{u}_\mathrm{s}^k,\boldsymbol{i}_\mathrm{s}^k)\Big), \\ \tfrac{\mathrm{d}}{\mathrm{d}t}i_\mathrm{s}^q &= \tfrac{1}{\widetilde{L}_s^q(\boldsymbol{i}_\mathrm{s}^k)}\Big(u_\mathrm{s}^q - R_\mathrm{s}i_\mathrm{s}^q + u_\mathrm{s,dist}^q(\omega_\mathrm{k},\boldsymbol{u}_\mathrm{s}^k,\boldsymbol{i}_\mathrm{s}^k)\Big). \end{aligned}\right\} \quad (13)$$

### B. Controller structure (see Fig. 3)

The proposed controller structure (following the idea in e.g. [11, Sec. 7.1.1], [20] or [3] with the same notation as here) consists of two parts, i.e.

$$\boldsymbol{u}_\mathrm{s,ref}^k = \underbrace{\boldsymbol{u}_\mathrm{s,pi}^k}_{\text{PI controller output}} + \underbrace{\boldsymbol{u}_\mathrm{s,comp}^k}_{\text{disturbance compensation}}. \quad (14)$$

Hence, the stator voltage reference $\boldsymbol{u}_\mathrm{s,ref}^k = (u_\mathrm{s,ref}^d, u_\mathrm{s,ref}^q)^\top$ – the control input to the VSI/electrical drive system – is the sum of the disturbance compensation $\boldsymbol{u}_\mathrm{s,comp}^k = (u_\mathrm{s,comp}^d, u_\mathrm{s,comp}^q)^\top$ and the output $\boldsymbol{u}_\mathrm{s,pi}^k = (u_\mathrm{s,pi}^d, u_\mathrm{s,pi}^q)^\top$ of the current PI controllers. Important to note that both controller parts will require online tracking of the flux linkages $\boldsymbol{\psi}_\mathrm{s}^k(\boldsymbol{i}_\mathrm{s}^k)$ (see Fig. 1) and of the differential inductances $L_\mathrm{s}^d(\boldsymbol{i}_\mathrm{s}^k)$, $L_\mathrm{s}^q(\boldsymbol{i}_\mathrm{s}^k)$ and $M(\boldsymbol{i}_\mathrm{s}^k) = L_\mathrm{s}^{qd}(\boldsymbol{i}_\mathrm{s}^k) = L_\mathrm{s}^{dq}(\boldsymbol{i}_\mathrm{s}^k)$ (see Fig. 2). Both parts will be discussed in more detail in the following.

### C. Disturbance compensation (feedforward control)

To obtain *decoupled* system dynamics in (13) for controller design, the "disturbance" $\boldsymbol{u}_\mathrm{s,dist}^k(t)$ as in (12) is compensated for by introducing the (ideal) feedforward compensation

$$\begin{aligned}\boldsymbol{u}_\mathrm{s,comp}^k(t) &:= \begin{pmatrix} u_\mathrm{s,comp}^d(t) \\ u_\mathrm{s,comp}^q(t)\end{pmatrix} = -\boldsymbol{u}_\mathrm{s,dist}^k(t + T_\mathrm{delay}) \\ \circ\!\!-\!\!\bullet\quad \boldsymbol{u}_\mathrm{s,comp}^k(s) &\approx \underbrace{\tfrac{c_0(1+sT_\mathrm{delay})}{1+sT_0}}_{=:F_\mathrm{comp}(s)}\boldsymbol{u}_\mathrm{s,dist}^k(s), \end{aligned} \quad (15)$$

where $0 < c_0 \leq 1$ and $0 < T_0 \ll T_\mathrm{delay}$. In view of (12), (15) requires knowledge of the resistance $R_\mathrm{s}$, the inductances $L_\mathrm{s}^d(\boldsymbol{i}_\mathrm{s}^k)$, $L_\mathrm{s}^q(\boldsymbol{i}_\mathrm{s}^k)$ and $M(\boldsymbol{i}_\mathrm{s}^k)$ (available as look-up table as shown in Fig. 2), the stator voltage $\boldsymbol{u}_\mathrm{s}^k$ (which can be approximated by the stator voltage reference $\boldsymbol{u}_\mathrm{s,ref}^k$), the stator current $\boldsymbol{i}_\mathrm{s}^k$ (measured), the stator flux linkage $\boldsymbol{\psi}_\mathrm{s}^k$ (available as look-up table as shown in Fig. 1) and the electrical angular velocity $\omega_\mathrm{k} = p\omega_\mathrm{m}$ ($\omega_\mathrm{m}$ measured). Actual flux linkages $\boldsymbol{\psi}_\mathrm{s}^k(\boldsymbol{i}_\mathrm{s}^k)$ and inductances $L_\mathrm{s}^d(\boldsymbol{i}_\mathrm{s}^k)$, $L_\mathrm{s}^q(\boldsymbol{i}_\mathrm{s}^k)$ and $M(\boldsymbol{i}_\mathrm{s}^k)$ are tracked online based on the actual current measurements.

Note that, in view of Assumptions 2 and 3, for $c_0 = 1$ and *small* $T_0 \ll T_\mathrm{delay}$, the compensator $F_\mathrm{comp}(s)$ in (15) approximates the inverse of the VSI dynamics (6) and allows to compensate for the disturbance $\boldsymbol{u}_\mathrm{s,dist}^k$ (almost) ideally by the feedforward term $\boldsymbol{u}_\mathrm{s,comp}^k$ (see [3]). So, nonlinear but *decoupled* currents dynamics are obtained as follows

$$\left.\begin{aligned} \tfrac{\mathrm{d}}{\mathrm{d}t}i_\mathrm{s}^d &= \tfrac{1}{\widetilde{L}_s^d(\boldsymbol{i}_\mathrm{s}^k)}\big(u_\mathrm{s}^d - R_\mathrm{s}i_\mathrm{s}^d\big), \\ \tfrac{\mathrm{d}}{\mathrm{d}t}i_\mathrm{s}^q &= \tfrac{1}{\widetilde{L}_s^q(\boldsymbol{i}_\mathrm{s}^k)}\big(u_\mathrm{s}^q - R_\mathrm{s}i_\mathrm{s}^q\big). \end{aligned}\right\} \quad (16)$$

**Remark 3** (Linear RSM). *For constant auxiliary inductances, i.e.* $\widetilde{L}_s^d(\boldsymbol{i}_\mathrm{s}^k) = \widetilde{L}_s^d > 0$ *and* $\widetilde{L}_s^q(\boldsymbol{i}_\mathrm{s}^k) = \widetilde{L}_s^q > 0$, *the current*

[4] I.e., $\boldsymbol{L}_\mathrm{s}^k(\boldsymbol{i}_\mathrm{s}^k)\boldsymbol{L}_\mathrm{s}^k(\boldsymbol{i}_\mathrm{s}^k)^{-1} = \boldsymbol{L}_\mathrm{s}^k(\boldsymbol{i}_\mathrm{s}^k)^{-1}\boldsymbol{L}_\mathrm{s}^k(\boldsymbol{i}_\mathrm{s}^k) = \boldsymbol{I}_2$ for all $\boldsymbol{i}_\mathrm{s}^k \in \mathbb{R}^2$.

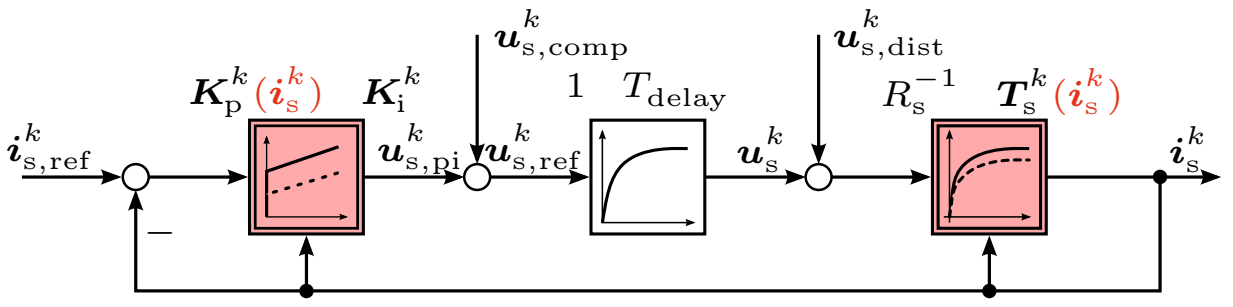

Fig. 3: Block diagram of the nonlinear current closed-loop system.

*dynamics* (16) *become linear decoupled first-order lag systems with the transfer functions*

$$\frac{i^d_{\mathrm{s}}(s)}{u^d_{\mathrm{s}}(s)} = \frac{\frac{1}{R_{\mathrm{s}}}}{1+sT^d_{\mathrm{s}}} \quad \wedge \quad \frac{i^q_{\mathrm{s}}(s)}{u^q_{\mathrm{s}}(s)} = \frac{\frac{1}{R_{\mathrm{s}}}}{1+sT^q_{\mathrm{s}}}, \tag{17}$$

*which share the same system gain* $1/R_{\mathrm{s}}$ *but have different time constants* $T^d_{\mathrm{s}} := \widetilde{L}^d_s/R_{\mathrm{s}}$ *and* $T^q_{\mathrm{s}} := \widetilde{L}^q_s/R_{\mathrm{s}}$.

### *D. Nonlinear PI controller design with anti-windup*

It is well known that PI(D) controllers in the presence of input saturation may exhibit integral windup (in particular for large initial errors) leading to large overshoots and/or oscillations in the closed-loop system response (see, e.g., [39, 40]). Therefore, a simple but effective anti-windup strategy (similar to 'conditional integration', see e.g. [40]) is implemented which stops integration of the integral control action if the control input (here $\boldsymbol{u}^k_{\mathrm{s}}$ or $\boldsymbol{u}^k_{\mathrm{s,ref}}$) exceeds the admissible range. For this, the "anti-windup decision function"

$$f_{\widehat{u}}(\boldsymbol{u}^k_{\mathrm{s,ref}}) := \begin{cases} 0, & \|\boldsymbol{u}^k_{\mathrm{s,ref}}\| \geq \widehat{u} \quad \left(\overset{(5)}{=} \frac{u_{\mathrm{dc}}}{2}\right), \\ 1, & \|\boldsymbol{u}^k_{\mathrm{s,ref}}\| < \widehat{u} \end{cases} \tag{18}$$

is combined with the nonlinear PI controller as follows

$$\left.\begin{aligned} \tfrac{\mathrm{d}}{\mathrm{d}t}\boldsymbol{\xi}^k_{\mathrm{i}} &= f_{\widehat{u}}(\boldsymbol{u}^k_{\mathrm{s,ref}})\, \boldsymbol{K}^k_{\mathrm{i}}\, (\boldsymbol{i}^k_{\mathrm{s,ref}} - \boldsymbol{i}^k_{\mathrm{s}}), \\ \boldsymbol{u}^k_{\mathrm{s,pi}} &= \boldsymbol{\xi}^k_{\mathrm{i}} + \boldsymbol{K}^k_{\mathrm{p}}(\boldsymbol{i}^k_{\mathrm{s}})\, (\boldsymbol{i}^k_{\mathrm{s,ref}} - \boldsymbol{i}^k_{\mathrm{s}}). \end{aligned}\right\} \tag{19}$$

where $\boldsymbol{\xi}^k_{\mathrm{i}} = (\xi^d_{\mathrm{i}}, \xi^q_{\mathrm{i}})^\top$ are the integrator outputs of the PI controllers. The controller gains (merged in the gain matrices)

$$\boldsymbol{K}^k_{\mathrm{p}}(\boldsymbol{i}^k_{\mathrm{s}}) := \begin{bmatrix} k^d_{\mathrm{p}}(\boldsymbol{i}^k_{\mathrm{s}}) & 0 \\ 0 & k^q_{\mathrm{p}}(\boldsymbol{i}^k_{\mathrm{s}}) \end{bmatrix} \wedge \boldsymbol{K}^k_{\mathrm{i}} := \begin{bmatrix} k^d_{\mathrm{i}} & 0 \\ 0 & k^q_{\mathrm{i}} \end{bmatrix} \tag{20}$$

are tuned according to the "Magnitude Optimum criterion" (see [1] or [11, p. 81,82]) as follows

$$\begin{aligned} d\text{-gains:} \quad k^d_{\mathrm{p}}(\boldsymbol{i}^k_{\mathrm{s}}) &= \frac{\widetilde{L}^d_s(\boldsymbol{i}^k_{\mathrm{s}})}{2T_{\mathrm{delay}}} > 0 \; [\Omega] \\ k^d_{\mathrm{i}} &= \frac{R_{\mathrm{s}}}{2T_{\mathrm{delay}}} > 0 \; [\mathrm{V}/(\mathrm{As})] \\ q\text{-gains:} \quad k^q_{\mathrm{p}}(\boldsymbol{i}^k_{\mathrm{s}}) &= \frac{\widetilde{L}^q_s(\boldsymbol{i}^k_{\mathrm{s}})}{2T_{\mathrm{delay}}} > 0 \; [\Omega] \\ k^q_{\mathrm{i}} &= \frac{R_{\mathrm{s}}}{2T_{\mathrm{delay}}} > 0 \; [\mathrm{V}/(\mathrm{As})]\,. \end{aligned} \tag{21}$$

Note that the integrator gains $k^d_{\mathrm{i}}$ and $k^q_{\mathrm{i}}$ are *constant*.

**Remark 4.** *The basic idea of the Magnitude Optimum is to compensate for the* large *time "constants"* $T^d_{\mathrm{s}}(\boldsymbol{i}^k_{\mathrm{s}}) := \widetilde{L}^d_s(\boldsymbol{i}^k_{\mathrm{s}})/R_{\mathrm{s}}$ *and* $T^q_{\mathrm{s}}(\boldsymbol{i}^k_{\mathrm{s}}) := \widetilde{L}^q_s(\boldsymbol{i}^k_{\mathrm{s}})/R_{\mathrm{s}}$ *of the system dynamics* (16) *[1]. Since,* $\widetilde{L}^d_s(\boldsymbol{i}^k_{\mathrm{s}})$ *and* $\widetilde{L}^q_s(\boldsymbol{i}^k_{\mathrm{s}})$ *are varying with the stator currents, an online adjustment of the controller parameters is required.*

**Remark 5.** *The use of the* discontinuous *anti-windup function as in* (18) *may lead to chattering [39]. Chattering was* not *observed in the simulations and experiments, however if chattering occurs, the use of a Lipschitz continuous anti-windup function, as proposed in [29, 41], might be beneficial.*

**Remark 6.** *If, for the current measurement, a first-order low-pass filter* $F_{\mathrm{f}}(s) = \frac{1}{1+s\,T_{\mathrm{f}}}$ *with filter time constant* $T_{\mathrm{f}} \ll 1\,\mathrm{s}$ *is implemented to filter out noise, the small time constants* $T_{\mathrm{delay}}$ *and* $T_{\mathrm{f}}$ *can be combined to one small time constant* $T'_{\mathrm{delay}} := T_{\mathrm{delay}} + T_{\mathrm{f}}$ *and, in* (21), $T'_{\mathrm{delay}}$ *should be used instead of* $T_{\mathrm{delay}}$ *[11, Sec. 7.1.1.5].*

## IV. Simulative and experimental validation

In this section, the proposed nonlinear PI controller (19) with anti-wind up (18) and online parameter adjustment (21) and disturbance compensation (15) with online tracking of the flux linkage and the differential inductances are implemented in Matlab/Simulink and at a laboratory setup. Key data of implementation, simulation and hardware is collected in Tab. I. Simulation and measurement results illustrate the effectiveness of the proposed controller structure with online parameter adjustment. For the *same* current reference trajectory, the control performance is illustrated for *two* scenarios over the whole operation regime of the RSM: (i) Operation at idle speed: —— simulation results and —— measurement results (see Fig. 6), and (ii) Operation with and without disturbance compensation (feedforward control): —— and —— measurement results (see Fig. 7). To ease comparability of the results, simulation and measurement results are both depicted in Figures 6 where the top large subfigure shows the overall duration of the experiment including the quantities (from top to bottom): direct currents $i^d_{\mathrm{s}}$ & $i^d_{\mathrm{s,ref}}$, quadrature currents $i^q_{\mathrm{s}}$ & $i^q_{\mathrm{s,ref}}$, norm of the control input $\|\boldsymbol{u}^k_{\mathrm{s,ref}}\|$ (& actuator limit $\widehat{u} = \frac{u_{\mathrm{dc}}}{2}$), and machine angular velocity $\omega_{\mathrm{m}}$. The three smaller subfigures at the bottom show different details (zoomed in excerpts) of the overall experiment including the quantities (from top to bottom): direct current $i^d_{\mathrm{s}}$ & $i^d_{\mathrm{s,ref}}$ and direct reference voltage $u^d_{\mathrm{s,ref}}$ (& actuator limit $\widehat{u} = \frac{u_{\mathrm{dc}}}{2}$), quadrature currents $i^q_{\mathrm{s}}$ & $i^q_{\mathrm{s,ref}}$ and quadrature reference voltage $u^q_{\mathrm{s,ref}}$ (& actuator limit $\widehat{u} = \frac{u_{\mathrm{dc}}}{2}$), and machine angular velocity $\omega_{\mathrm{m}}$.

### *A. Implementation in Matlab/Simulink*

The implementation in Matlab/Simulink is shown in Fig. 4. Machine model (1) and VSI model (3) with $\boldsymbol{u}^{ltl}_{\mathrm{s}}$ as in (2) and constant DC-link voltage are implemented in the $s = (\alpha, \beta)$-reference frame. The implemented regularly sampled, symmetrical PWM (for details see [32, Sec. 8.4.6]) applies the required pulse pattern (switching vector $\boldsymbol{s}^{abc}_{\mathrm{s}}$) according to the desired reference stator voltage $\boldsymbol{u}^k_{\mathrm{s,ref}}$. The implementation of the nonlinear PI controller (19) and the disturbance compensation (15) is in the $k = (d, q)$-reference frame. The overall implementation is depicted as block diagram in Fig. 4. The feedforward control (15) is implemented with static compensator, i.e., $F_{\mathrm{comp}}(s) = 1$, and the use of $\boldsymbol{u}^k_{\mathrm{s,ref}}$ instead of the stator voltages $\boldsymbol{u}^k_{\mathrm{s}}$ (which are not measured in real world and hence are not available for feedback). For the simulation, an inverse interpolation algorithm is used to extract the actual stator currents $\boldsymbol{i}_s^s = \boldsymbol{T}_p(\phi_k)\boldsymbol{i}_s^k$ from the available flux maps.

| description | symbols & values [unit] |
|---|---|
| RSM (9) | $p_{\text{mech}} = 1{,}1\,\text{kW}$, $R_s = 4{,}7\,\Omega$, $p = 2$, $\boldsymbol{\psi}_s^k$ as in Fig. 1, $\|\boldsymbol{i}_s^k\| \leq 5\,\text{A}$, $n_{\text{nominal}} = 1\,500\,\text{rpm}$, $I_{\text{nominal}}^{\text{rms}} \approx 2\,\text{A}$, $U_{\text{nominal}}^{\text{rms}} = 400\,\text{V}$ (@50 Hz) |
| Mechanics | $\Theta = 8{,}1 \cdot 10^{-3}\,\text{kg}\,\text{m}^2$ (friction neglected) |
| VSI | $u_{\text{dc}} = 600\,\text{V}$, $f_s = 5\,\text{kHz}$, $T_{\text{delay}} = \frac{3}{2f_s}$ |
| RT-system | $T_s = 1/f_s$ (sampling period), measured signals were *not* filtered! |
| reference | see 1st & 2nd subplots in Fig. 6 |
| disturbance comp. (15) | $F_{\text{comp}}(s) = 1$ (VSI dynamics neglected), In (12), instead of $\boldsymbol{u}_s^k$, $\boldsymbol{u}_{s,\text{ref}}^k$ is used |
| anti-windup PI (19), (18) | $\widehat{u} = u_{\text{dc}}/2 = 300\,\text{V}$ (due to symmetrical, regularly sampled PWM), $\boldsymbol{\xi}_i^{k,0} = (0,0)^\top$ |

Table I: System, implementation and controller data.

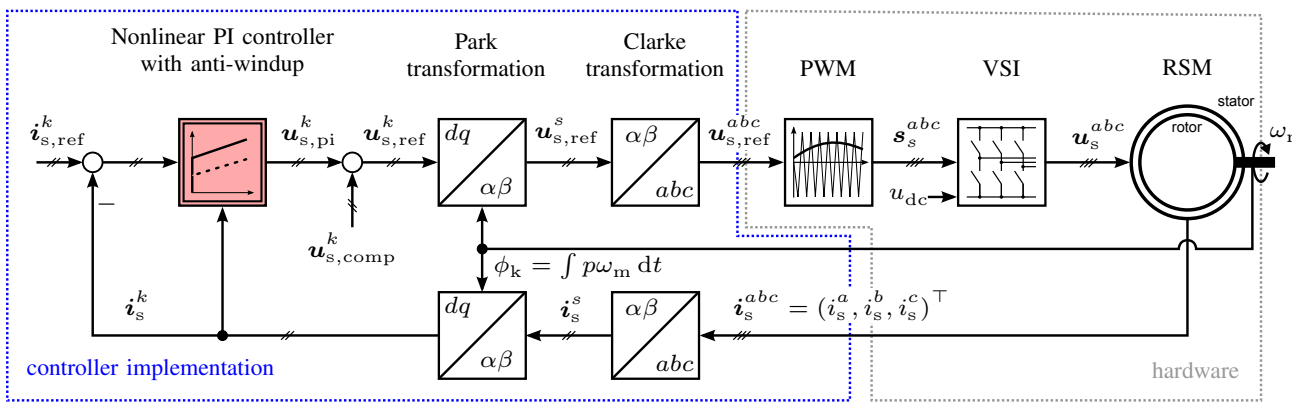

Fig. 4: Block diagram of the implemented simulation model.

### *B. Implementation at the laboratory setup*

*1) Laboratory setup:* The laboratory setup is depicted in Fig. 5. It consists of the RSM (on the left) under test (with flux linkages as shown in Fig. 1) and an induction machine (on the right; not controlled for the experiments). Both machines are driven by voltage source inverters of the same power rating. The implementation is done in C on a Pentium-based real-time system. To illustrate and validate the control performance as close as possible with respect to the proposed theoretical controller design presented in Sec. III, **no filters** were implemented for current measurement. Hence, the measured signals in Figures 6 and 7 exhibit noticable noise. However, the control performance is still acceptable, fast, and accurate over the whole operation range of the RSM.

*2) Discrete-time implementation at the real-time system:* For the discrete implementation, the nonlinear PI controller (19) with anti-windup and the disturbance compensation (15) are discretized by the (explicit) Euler method (i.e., $\frac{\mathrm{d}}{\mathrm{d}t}x(t) \approx \frac{x[k+1]-x[k]}{T_s}$ and $x(t) \approx x[k]$ where $x[k] := x(kT_s)$ and $T_s = \frac{1}{f_s} \ll 1\,\text{s}$) and implemented in C.

### *C. Discussion of the results*

*1) Discussion of Scenario (i) — Idle speed (see Fig. 6):* The simulation results are shown in cyan (see —— in Fig. 6) whereas the measurements results are shown in blue (see —— in Fig. 6). The control performance of the proposed control strategy for current reference tracking is fast and almost perfectly accurate. Solely, for higher currents (higher torque), the machine is rotating faster which leads to higher and ramp-like changes of the machine speed and hence to higher and ramp-like changes of the back EMF. In turn, the controller structure generates a higher and ramp-like stator voltage reference to compensate for the disturbance (12). The noise in the reference voltages $u_{s,\text{ref}}^d$ and $u_{s,\text{ref}}^q$ are due to the numerical (inverse) interpolation of the inductances and flux linkage values which was based on a look-up table with only 25x25 supporting points. Nevertheless, on average, the simulation results very much coincide with the measurement results. The remaining deviations are due to iron losses and hysteresis effects in the machine which were not considered in the machine model.

The measured control performance of the proposed control strategy for current reference tracking is very similar to the simulation results. It is fast and almost perfectly accurate. Again, due to higher and more rapid changing torques for higher currents, the machine speed exhibits faster and ramp-like changes which yields higher and ramp-like changes of the back EMF. The controller must counteract by applying higher and ramp-like stator voltage references.

Most important, in simulations and measurements, the closed-loop current dynamics show an (almost) identical transient behavior: Over the complete operation range with $\|\boldsymbol{i}_s^k\| \leq 5\,\text{A}$, rise and settling times are similar for direct and quadrature currents, respectively. Hence, outer control loops (e.g. for speed and/or position control) can be designed based on simple dynamical approximations of the underlying closed-loop current dynamics.

*2) Discussion of Scenario (ii)—The positive effect of using the disturbance compensation (see Fig. 7):* The measurement results *with* —— disturbance compensation (15) are shown in blue, whereas the measurement results *without* —— disturbance compensation (i.e., $\boldsymbol{u}_{s,\text{comp}}^k(t) = (0,0)^\top$) are shown in green. The results clearly illustrate the advantageous effect of the disturbance compensation on the control and tracking performance. *With* —— disturbance compensation, direct and quadrature currents are decoupled and track the reference steps accurately, whereas *without* —— disturbance compensation the tracking accuracy is deteriorated (in particular the tracking accuracy of the quadrature current).

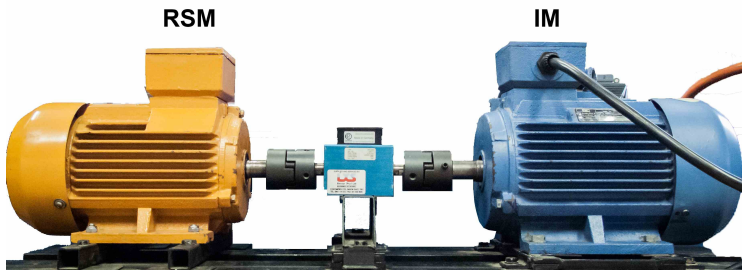

Fig. 5: Reluctance synchronous machine (RSM) and induction machine (IM).

## V. Conclusion and Future Work

In this paper, nonlinear PI controllers with anti-windup and online parameter adjustment have been proposed and implemented for current control of nonlinear reluctance synchronous machines (RSMs) where the flux linkage maps are known and available to the control engineer (e.g. in the form of look-up tables). The online adjustment of the PI controller parameters requires the online tracking of the differential inductances, which are approximated by numerical differentiation of the flux linkage maps with respect to the stator currents. In addition, a nonlinear disturbance compensation (feedforward control) with online tracking of the flux linkages and the differential inductances has been proposed and implemented to compensate for the nonlinear cross-coupling between the direct and quadrature current dynamics. The nonlinear PI controller design has been derived in state space and has been tuned according to Magnitude Optimum. The proposed controller structure has been implemented in Matlab/Simulink and at a laboratory test bench. Simulation and measurement results fit closely, and illustrate and validate the expected control performance of the proposed nonlinear control strategy. The closed-loop current dynamics exhibit (almost) identical rise and settling times over the whole operation range of the machine.

Future work will focus on (i) the consideration of iron losses (e.g. due to hysteresis and eddy currents), (ii) a robustness analysis (e.g. when the flux linkages are only roughly known), (iii) an analytical or numerical method for maximum-torque-per-Ampere reference current generation, and (iv) the combination with sensorless control methods.

## VI. Acknowledgment

The authors are deeply indebted to Dr.-Ing. Peter Landsmann (Technische Universität München) for very fruitful discussions and the interpolation algorithms for the implementation in Matlab/Simulink and C. This study and the mutual research visits in Germany and South Africa were made possible by the financial support granted by the "Freunde der Technischen Universität München e.V".

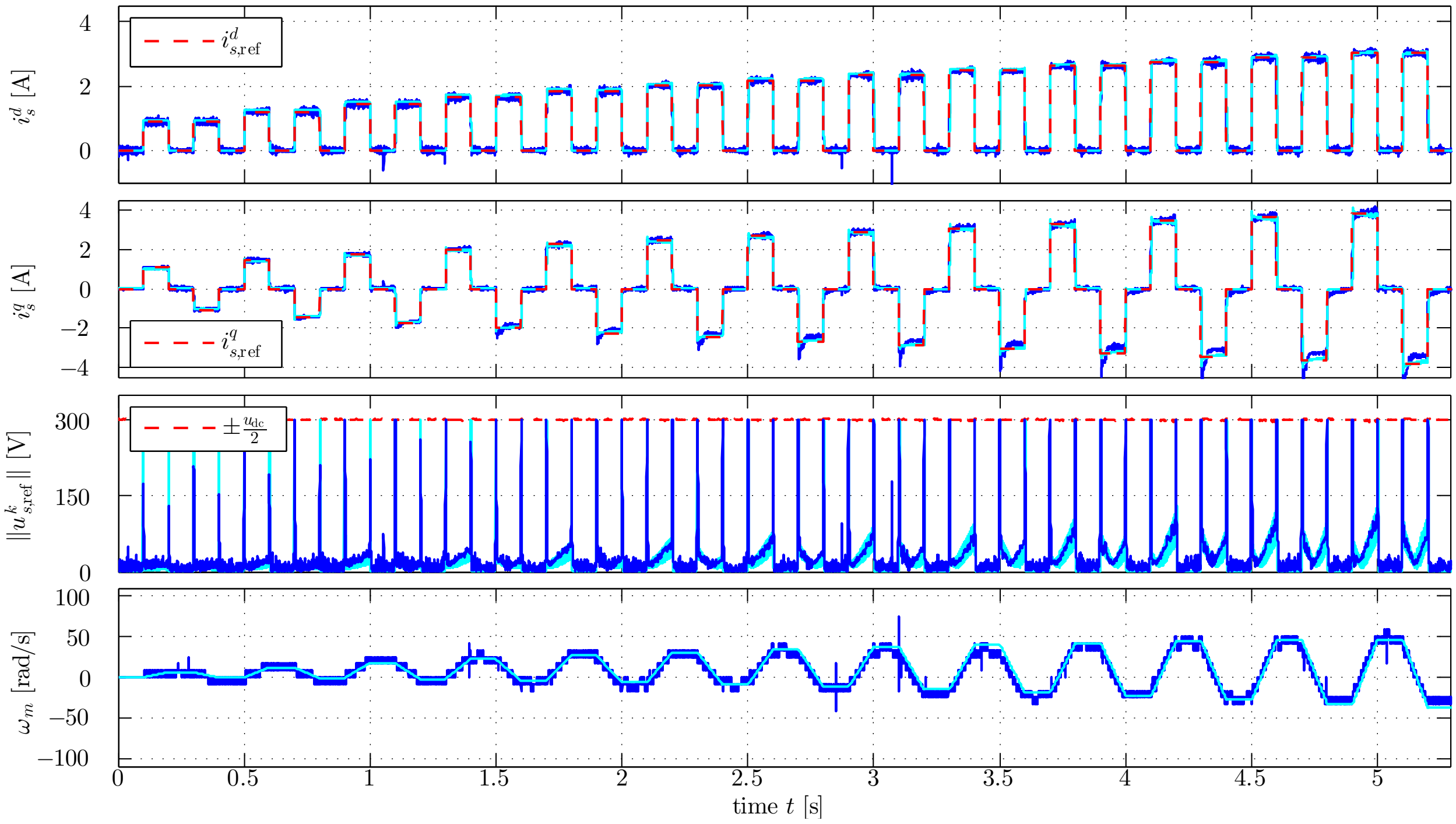

(a) Overall control performance (from top to bottom): currents $i_\mathrm{s}^d$ and $i_\mathrm{s}^q$, norm reference voltage $\|\boldsymbol{u}_{\mathrm{s,ref}}^k\|$, and angular velocity $\omega_\mathrm{m}$.

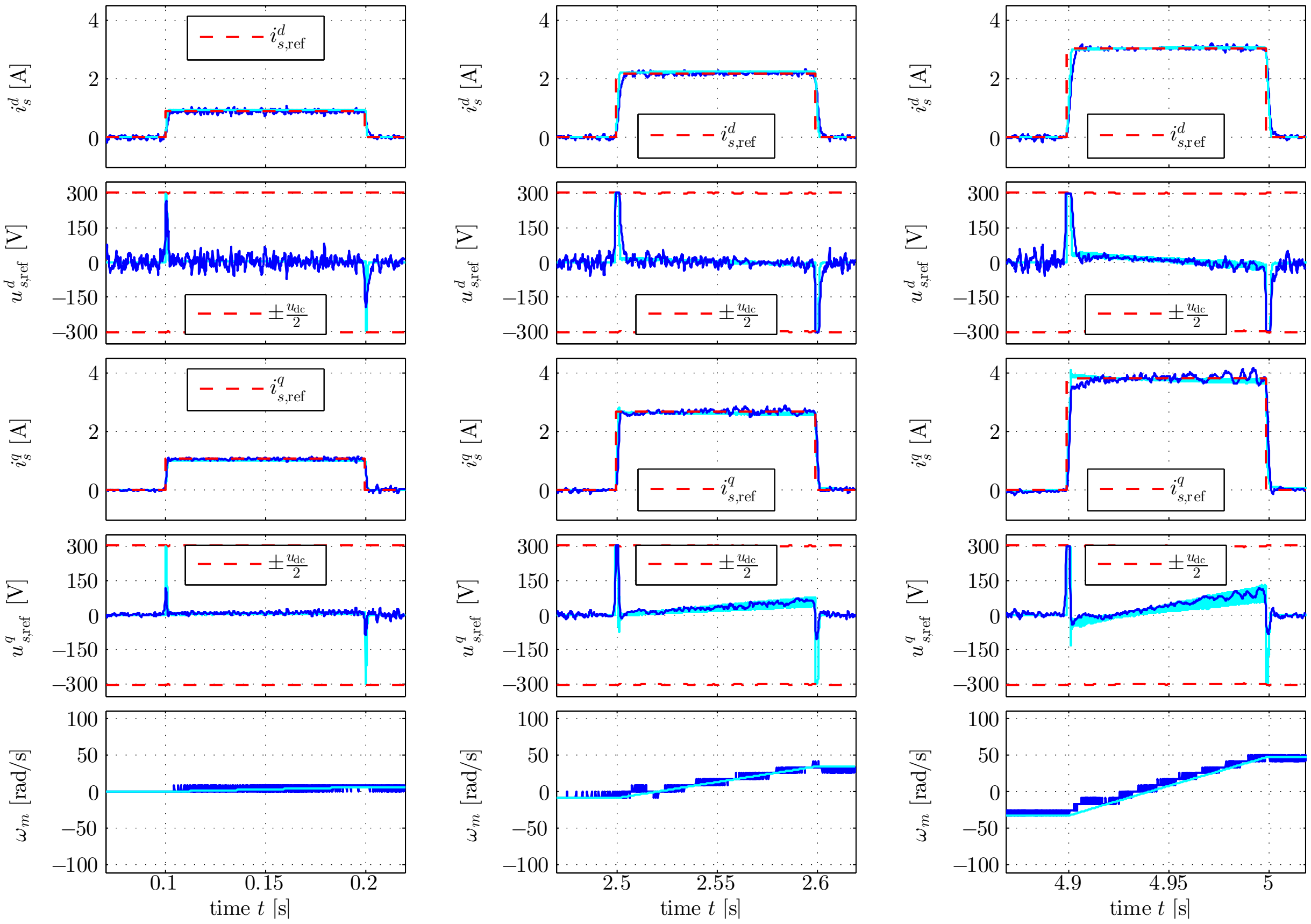

(b) Zoom 1.

(c) Zoom 2.

(d) Zoom 3.

Fig. 6: **Simulation results** *and* **Measurement results** at *idle speed* (**unfiltered!**) for the RSM with flux linkage as depicted in Fig. 1: Control performance of the proposed nonlinear PI current controller (19) with online parameter adjustment (21).

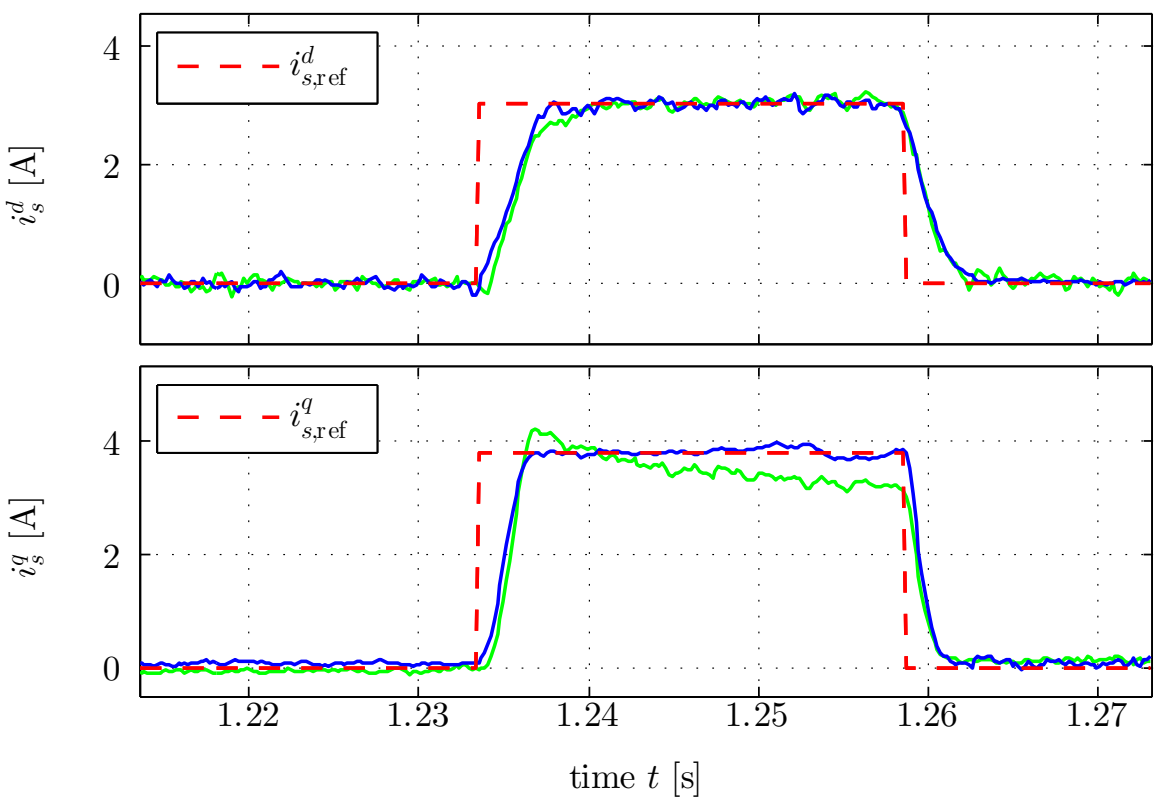

Fig. 7: **Measurement results** for the RSM with flux linkage as depicted in Fig. 1: Tracking performance (i) —— *with* disturbance compensation, i.e. $\boldsymbol{u}^k_{s,\mathrm{comp}} = -\boldsymbol{u}^k_{s,\mathrm{dist}} = -(u^d_{s,\mathrm{dist}}, u^q_{s,\mathrm{dist}})^\top$ (as in (15)) and (ii) —— *without* disturbance compensation, i.e. $\boldsymbol{u}^k_{s,\mathrm{comp}} = \boldsymbol{0}_2$.